\documentclass{camera}
\def\ref{\noindent\hangindent=20pt\hangafter=1} %istr. per bibliografia
\begin{document}

%%%%%%%%%%%%%%%%%%%%%%%%%%%%%%%%%%%%%%%%%%%%%%%%%%%%%%%%
% The title, all uppercase; if you want to split it in
% two or more lines, put a \\ macro at each line break
% example: 
%   \title{TITLE: FIRST LINE\\ SECOND LINE}
%
\title{OPTICAL OBSERVATIONS OF THE FIELD OF GRB 970111}
 
%%%%%%%%%%%%%%%%%%%%%%%%%%%%%%%%%%%%%%%%%%%%%%%%%%%%%%%%
% The author(s), separated by commas; do not put a
% comma before the last author, use instead the \And
% macro which produces a normal ``and'' in the
% caps/small caps context
%
\author{Nicola Masetti, Corrado Bartolini, Adriano Guarnieri \& 
Adalberto Piccioni}

%%%%%%%%%%%%%%%%%%%%%%%%%%%%%%%%%%%%%%%%%%%%%%%%%%%%%%%%
%
\organization{Dipartimento di Astronomia, Universit\`a di Bologna}

\maketitle

%%%%%%%%%%%%%%%%%%%%%%%%%%%%%%%%%%%%%%%%%%%%%%%%%%%%%%%%
% Write the text starting from here and using the usual
% LaTeX commands.
%
\par\noindent\centerline {\bf Abstract}

$B$, $V$ and $R$ optical photometry performed on the field of the GRB 970111 
soon after its detection by BeppoSAX is presented. No remarkable optical object
is found inside the corrected error box. The optical counterpart of an X--ray 
and radio source and a suspected red variable (probably an eclipsing binary)
have been detected inside the field of GRB 970111.

\section 
{\bf Introduction}

GRB 970111 was detected as a three--peaked Gamma--Ray Burst (hereafter GRB)
on January 11,
1997 by the Wide Field Camera of the X--ray satellite BeppoSAX (Costa et al. 
1997). This was the second GRB event (after GRB 960720; Piro et al. 1996) ever 
observed by BeppoSAX; the coordinates of the burst were: 
$\alpha=15^{\rm h} 28^{\rm m} 24^{\rm s}$, $\delta=19^{\circ} 40'.0$ 
(equinox 2000.0; error box radius = 10'). Soon after, 
Butler et al. (1997) reported that two faint X--ray sources, hereafter 
labelled as `a' and `b', are present in the field of GRB 970111.
According to Guarnieri et al. (1997a), no object has shown
remarkable variations in the optical, neither in the field
of GRB 970111 nor in the error boxes of the two X--ray sources.
Then, Hurley et al. (1997a) reduced the error box of GRB 970111 and found that 
only BeppoSAX source `a' lies within it. Source `a' appears to be also an
optical (Guarnieri et al. 1997b, Kulkarni et al., 1997) and radio source 
(Frail et al. 1997a).
This error box was one of the smallest ever given for a GRB, making it suitable
for the search for optical counterparts. The width of the error box did 
not allow us to image it in one single frame, so we focused our attention on 
the X--ray sources, in particular on source `a'.
Unfortunately, due to a misalignment of the Wide Field Cameras of BeppoSAX 
(in't Zand et al. 1997), the error box reported by Costa et al. (1997) had to 
be shifted $\sim$3' from its previous position; then, Hurley et al. (1997b) 
gave a new error box, seven times smaller than the former and located in its 
southern part.

\par~
In this paper we present an optical photometric analysis mainly concerning the 
northern and central parts of the GRB 970111 former error box (Hurley et al. 
1997a),
observed from January 14 (i.e. $\sim$65 hours after the burst) to March 12. 
Section 2 will briefly describe the reduction and calibration techniques,
while Section 3 will present and analyse the results.

\section 
{\bf The observations} 

The frames have been obtained with the 1.5-meter telescope (equipped with the 
BFOSC instrumentation) of the Bologna Astronomical Observatory on January 14, 
15, 17 and 31, on February 14, 17 and 18, and on March 5 and 12, 1997.
$B$, $V$ and $R$ filters were employed. We used long (15 to 75 minutes) 
exposure times to reach limiting magnitudes ranging from $\sim$20.5 to $\sim$22.
$B$, $V$ and $R$ images of the PG 1047+003 sequence (Landolt 1992) were also 
obtained in order to calibrate the field of GRB 970111.

A comparison among frames acquired on the same night in different bands
and performed with a simple FORTRAN code allowed us to determine the $B-V$, 
$V-R$ and/or $B-R$ colors of most stars within the former GRB 970111 field.
The same procedure was applied to frames collected on different nights
but in the same filter in order to search for the variable stars of the field.
This double cross--correlation allowed the selection of interesting object
which will be presented and discussed in the next Section.

\section 
{\bf Data analysis and discussion}

No object inside the various fields showed any remarkable blue excess (we found 
that no $B-V$ was lower than 0.5 and no $V-R$ was lower than 0.2). 
These fields are at fairly high galactic latitude ($b^{\rm II}=53^{\circ}$);
so, the interstellar absorption, and thus the $B-V$ color excess, should not be 
high. Thus, the values of the color indices quoted above should not be 
very different from the unabsorbed ones.
Therefore, the prime selection rule has been the light variability, with 
particular attention to the objects, inside the part of the corrected error box
contained in our images of January 14 and 17, which showed a decline in their 
luminosity. None of them presented a decrease of more than 0.3
mag in $R$, thus confirming the results by Castro--Tirado et al. (1997).

A frame containing the whole GRB 970111 corrected error box (Fig. 1a), taken 
on March 5 under photometric sky conditions, shows no object down to $R\sim22$ 
at the position of the radio source reported by Galama et al. (1997).
Actually, we see that this field is rich in galaxies, as we found about 30--40 
galaxies with $R<21$ in the GRB 970111 error box reported by Costa et al. 
(1997).

Two objects of the northern part of the extended error box (Hurley et al. 
1997a) have particularly drawn our attention: the optical counterpart (object 
1 in Fig. 1b) of ROSAT X--ray source `a' (Butler et al. 1997) and of the radio 
source VLA J1528.7+1945 (Frail et al. 1997a) and a long--term red variable 
(object 2 in Fig. 1b).

\par~

As regards object 1, it is almost certainly the optical counterpart of BeppoSAX
source `a' and of radio source VLA 1528.7+1945. Its position has been computed 
by means of the Digitized Sky Survey (hereafter DSS) and its coordinates 
coincide, within the errors, with the position of the radio source given by 
Frail et al. (1997a).
Object 1 appears as a $R=20.6\pm0.1$ object, while it is barely visible on the 
$V$ frame of January 31. It is instead invisible in the $B$ frame of the same 
night ($B>21$). On the Palomar plates, this object is undetectable in $B$ and 
is about at the same $R$ luminosity level stated above. On February 18, its $B$ 
and $V$ magnitudes were $21.7\pm0.1$ and $21.0\pm0.1$, respectively, while its 
$R$ magnitude was at about the same value of the January observations. This 
implies that its colors are $B-V=0.7$ and $V-R=0.4$. The March data confirm 
these figures.

The object has been resolved by Kulkarni et al. (1997) into two moderately 
redshifted galaxies separated by $\sim$2".5. Indeed, it is asymmetrical and 
seems to be formed by at least two (but probably four; see Fig. 1b) objects, 
with the two brighter ones coinciding with the galaxies S1 and S2 reported by 
Kulkarni et al. (1997).

\par~\par~\par~\par~\par~\par~\par~\par~\par~\par~\par~\par~
\par~\par~\par~\par~\par~\par~\par~\par~\par~

{\bf Fig. 1.} {\bf a} Corrected error box of GRB 970111 (Hurley et al. 1997b). 
The field, obtained in the $R$ band on March 5, covers an area of 
9'$\times$9' and has a limiting magnitude of $\sim$22.
{\bf b} Northern part of the former error box of GRB 970111 as indicated by 
Hurley et al. (1997a). This $R$ frame (area: 4'.5$\times$4'.5), obtained also 
on March 5, has the same orientation and limiting magnitude of that in Fig. 1a.
See text for further details on objects 1 and 2 indicated here.  

\par~\par~

Concerning the other interesting object, i.e. object 2, after a quick--look 
comparison between the field and the DSS, we noticed on our $R$ frames the 
clear presence of a star which was barely visible on the DSS (whose limiting 
magnitude is $R\sim20.5$). The star is also practically invisible in the 
Palomar Sky Survey red plates and absent in the blue ones (limiting magnitude 
$\sim$21 for both). From the DSS we deduced the coordinates of this object: 
$\alpha=15^{\rm h}$ $28^{\rm m}$ $45^{\rm s}$; $\delta=+19^{\circ}$ 47' 15" 
(equinox 2000.0), with a conservative error of $\pm$5" for both values. 
No variable object within a circle of radius 10' and centered on these 
coordinates is mentioned in the SIMBAD database.

The star, on January 15, 1997, was at $R=19.90\pm0.05$, thus showing a 
variation of more than a magnitude with respect to the Palomar red plates
(April 1950), while on January 17 the $R$ magnitude was $19.93\pm0.05$.
On January 31 its magnitude in the $R$ band decreased and was found to be 
$19.80\pm0.05$. During the same night, its $V-R$ color index was 0.54
(see also Masetti et al. 1997).

The February observations reveal that this object has a $B=21.8\pm0.1$
and a $V$ magnitude in agreement with the values of January 31. Its $B-V$ color
index is therefore 1.4. The colors suggest that this object is a mid--late K 
spectral type star, depending on the luminosity class (Lang 1992). 
It is noteworthy that, on the night of February 18, the star underwent 
a fading in the $R$ band of about 0.5 mag which lasted less than 3 hours,
possibly due to an eclipse. The March observations show this object at the 
$B$ and $R$ values outside the fading phase. The magnitudes obtained 
from the Palomar Sky Survey and from our frames then seem to indicate that 
this star might be a binary red dwarf which undergoes eclipses.

\par~

Finally, we noticed that in the DSS frame an extended feature is present at
$\alpha=15^{\rm h}$ $28^{\rm m}$ $32^{\rm s}$; $\delta=+19^{\circ}$ 39' 09"
($\pm$5"; equinox 2000.0).
This feature is missing both in all our frames and in $B$ and $R$ Palomar Sky
Survey plates. We suggest that this `ghost object' might be an error in the
digitalization of the Palomar plates.

\par~
\par~

{\it Acknowledgements}. This research made use of the SIMBAD data base, 
operated by Centre de Donn\'ees Stellaires in Strasbourg, France. We thank
M. Corwin and M. Salvato for useful comments and suggestions.

\section*
{\bf References}

\noindent
Butler R.C., Piro L., Costa E. et al., 1997, IAU Circ. 6539

\noindent
Castro--Tirado A.J., Gorosabel J., Heidt J. et al., 1997, IAU Circ. 6598

\noindent
Costa E., Feroci M., Piro L. et al., 1997, IAU Circ. 6533

\noindent
Frail D.A., Kulkarni S.R., Nicastro L. et al., 1997, IAU Circ. 6545

\noindent
Galama T., Strom R., van Paradijs J. et al., 1997, IAU Circ. 6571

\noindent
Guarnieri A., Bartolini C., Piccioni A. et al., 1997a, IAU Circ. 6544

\noindent
Guarnieri A., Bartolini C., Piccioni A., Masetti N., 1997b, IAU Circ. 6559

\noindent
Hurley K., Kouveliotou C., Fishman G., Meegan C., 1997a, IAU Circ. 6545

\noindent
Hurley K., Kouveliotou C., Fishman G. et al., 1997b, IAU Circ. 6571

\noindent
in't Zand J., Heise J., Hoyng P. et al., 1997, IAU Circ. 6569

\noindent
Kulkarni S.R., Metzger M.R., Frail D.A., 1997, IAU Circ. 6559

\noindent
Landolt A.U., 1992, AJ, 104, 340

\noindent
Lang K.R., 1992, Astrophysical Data: Planets and Stars. Springer--Verlag, New 
	York

\noindent
Masetti N., Bartolini C., Guarnieri A., Piccioni A., 1997, IBVS No. 4440

\noindent
Piro L., Costa E., Feroci M. et al., 1996, IAU Circ. 6467

%%%%%%%%%%%%%%%%%%%%%%%%%%%%%%%%%%%%%%%%%%%%%%%%%%%%%%%%
% End of the paper
%
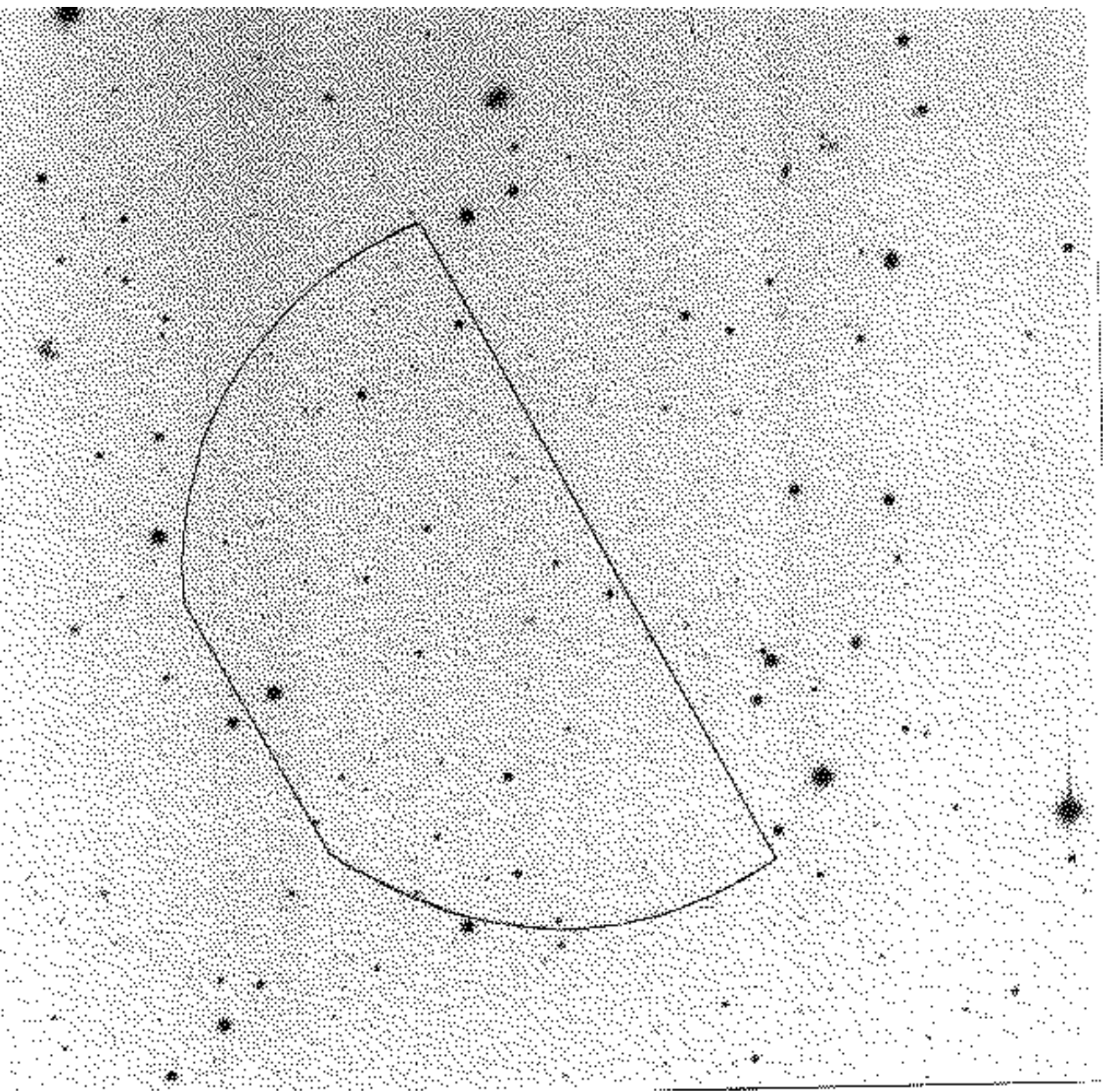
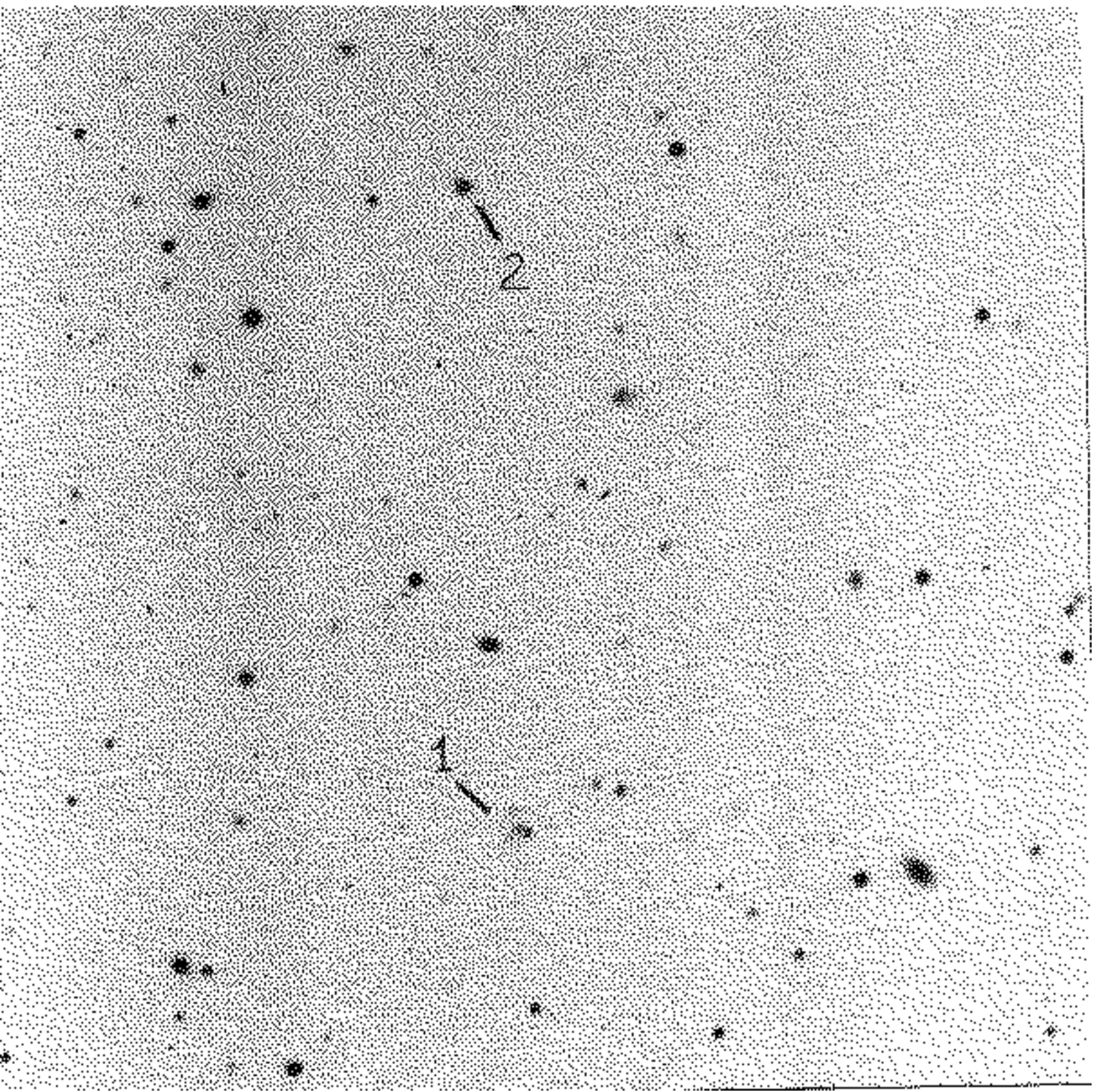
\end{document}